\documentclass[english,aip,apl,preprint,amsmath,amssymb,floatfix]{revtex4-1}
\usepackage[T1]{fontenc}
\usepackage[latin9]{inputenc}
\usepackage{textcomp}
\usepackage{graphicx}

\makeatletter
\@ifundefined{textcolor}{}
{%
 \definecolor{BLACK}{gray}{0}
 \definecolor{WHITE}{gray}{1}
 \definecolor{RED}{rgb}{1,0,0}
 \definecolor{GREEN}{rgb}{0,1,0}
 \definecolor{BLUE}{rgb}{0,0,1}
 \definecolor{CYAN}{cmyk}{1,0,0,0}
 \definecolor{MAGENTA}{cmyk}{0,1,0,0}
 \definecolor{YELLOW}{cmyk}{0,0,1,0}
 }

\makeatother

\usepackage{babel}
\begin{document}

\preprint{This line only printed with preprint option}

\title{Local Charge Transfer Doping in Suspended Graphene Nanojunctions}

\author{Jeffrey H. Worne}

\affiliation{Department of Electrical and Computer Engineering, Rice University,
6100 Main Street, Houston, Texas 77005, USA}

\author{Hemtej Gullapalli}
\author{Charudatta Galande}
\author{Pulickel M. Ajayan}

\affiliation{Department of Mechanical Engineering and Materials Science, Rice
University, 6100 Main Street, Houston, Texas 77005, USA}



\author{Douglas Natelson}

\email{natelson@rice.edu}

\affiliation{Department of Physics and Astronomy, Rice University, 6100 Main Street,
Houston, Texas 77005, USA}

\affiliation{Department of Electrical and Computer Engineering, Rice University,
6100 Main Street, Houston, Texas 77005, USA}

\begin{abstract}
We report electronic transport measurements in nanoscale graphene
transistors with gold and platinum electrodes whose channel lengths
are shorter than 100 nm, and compare them with transistors with channel
lengths from 1 \textmu{}m to 50 \textmu{}m. We find a large positive
gate voltage shift in charge neutrality point (NP) for transistors
made with platinum electrodes but negligible shift for devices made
with gold electrodes. This is consistent with the transfer of electrons
from graphene into the platinum electrodes. As the channel length
increases, the disparity between the measured NP using gold and platinum
electrodes disappears. 
\end{abstract}
\maketitle

Understanding the band alignment between metals and graphene is important
in graphene-based field-effect transistors,\cite{Liao2010} photovoltaic
devices,\cite{Mueller2010} and other experimental graphitic systems.\cite{Geim2009a}
Charge transfer is often seen as undesirable, for example between
a Si/SiO$_{2}$ substrate and graphene.\cite{Novoselov2007,Du2008}
Selective doping using reactive metal ions\cite{Chen2008a} or chemical
species\cite{Wang2009,Wehling2008} can tailor graphene's electrical
properties, however, effectively doping it with holes or electrons.
Recent work on the metal-graphene interface has focused on scanning
photocurrent measurements\cite{Lee2008,Mueller2009,Sundaram2009}
while electrical measurements using 4-probe geometries have examined
the influence of metal contacts on charge transport.\cite{Huard2008,Blake2009}

The role of metal contacts in graphene devices has only recently begun
to be explored. The interaction between rolled sheets of graphene
$-$ nanotubes $-$ and metals has been well-studied.\cite{Rao1997,Petit1999,Lang2000}
The unique band structure found in graphene,\cite{Zhang2005} however,
requires a different treatment. Charge carriers in graphene behave
like massless Dirac fermions.\cite{Novoselov2005} As such, they cause
suppressed charge screening and, when in contact with a metal, lead
to a long-range inhomogeneous electrostatic potential\cite{Khomyakov2010}
that can extend for several hundred nanometers into the bulk.\cite{Mueller2009,Golizadeh-Mojarad2009}
This can lead to marked changes in charge transport,\cite{Giovannetti2008,Golizadeh-Mojarad2009}
p-n junctions at the metal-graphene interface,\cite{Khomyakov2009,Mueller2010}
charge density fluctuations near the contacts,\cite{Jalilian2011}
and, for particular metals, the formation of a band gap.\cite{Varykhalov2010}

Due to the length scale of charge diffusion, probing charge transfer
is possible provided the electrode separation is no greater than twice
the diffusion distance for charge. Experimentally, this should manifest
itself as a shift in the charge neutrality point (NP) as the metal
contacts effectively make the local graphene environment n- or p-type.
Using a technique developed previously,\cite{Fursina2008} we are
able to fabricate 3-terminal suspended single-layer graphene transistors
with a high aspect ratio (width/length $\sim300$) and channel lengths less
than 100 nm. With transport measurements, we measure shifts in the
NP in graphene, suggesting that we are measuring regions of increased
charge density over the bulk.

\begin{figure}[p]
\includegraphics[width=8cm, clip]{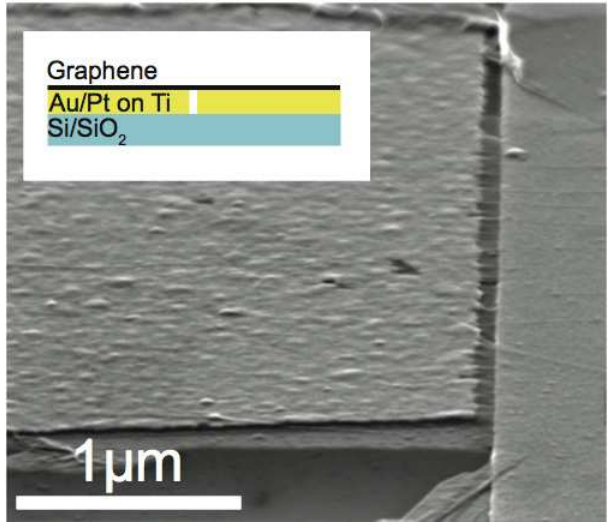}

\caption{\label{fig:fig1}SEM micrograph of graphene on gold electrodes. The
electrode separation is 72~nm. Wrinkles and folds can be followed across
the gap, suggesting the graphene is suspended. Inset: Cartoon illustrating
sample geometry.}

\end{figure}

We used degenerately doped p-type silicon with 200 nm of thermally
grown oxide, which serves both as the substrate and the gate. Using
two-step electron-beam lithography, electron-beam evaporation, and
lift-off processing, electrodes with $w=20$ \textmu{}m were defined
based on the previously mentioned technique with an average channel
length of 51 nm. These short channel electrodes are composed of a
1 nm titanium adhesion layer and 15 nm of gold or platinum. For comparison,
a set of interdigitated electrodes with $l=1-50$ \textmu{}m and $w=200$
\textmu{}m were fabricated with gold or platinum as the electrode.
Graphene was grown on copper foils and transferred to our substrate/electrodes
using poly(methyl methacrylate) (PMMA) as a transfer medium.\cite{Li2009}
The quality of the graphene was determined using Raman spectroscopy
and was verified to be single layer. A representative device is shown
in Figure \ref{fig:fig1}. 

Samples were measured in a variable temperature probe station and
data from 300K to 4K was collected using a HP4145A parameter analyzer.
Mobilities were calculated\cite{Chen2008a} by evaluating $d\sigma/dV_{g}=\mu C_{g}$
at the largest value of $d\sigma/dV_{g}$. $C_{g}$ is the capacitance
per unit area of the gate, and, since our graphene is suspended, a
serial capacitor model is used incorporating contributions from the
oxide and vacuum under the graphene giving $C_{g}=7.26\times10^{-8}$
F cm$^{-2}$. For both gold and platinum electrodes, the highest values
of mobilities were found at 20K.  Note that bias-driven self-heating may be an issue
at the lowest temperatures examined\cite{Checkelsky2008}, though this does not
affect the main observations of this work.  For the short channel devices made
with gold, $\mu_{20K}=4.8\times10^{2}$ cm$^{2}$ / V $\cdot$ s and
for short channel devices with platinum electrodes, $\mu_{20K}=3\times10^{1}$
cm$^{2}$ / V $\cdot$ s. The mobility for the interdigitated electrode
devices was $3.0\times10^{3}$ cm$^{2}$ / V $\cdot$ s and $1.0\times10^{3}$
cm$^{2}$ / V $\cdot$ s for the gold and platinum electrodes, respectively.
We note that the measured values of mobility are well below those
traditionally quoted in the literature of $\sim4000$ for graphene
fabricated under similar growth and transfer conditions.\cite{Li2009}
This is likely caused by defects introduced during the growth and
transfer process or by structural irregularities such as the folds
and ripples visible in Figure \ref{fig:fig1}. Ripples have been found
experimentally to increase resistivity in graphene and limit the mobility
of charge carriers.\cite{Katsnelson2008} The difference in mobilities
between samples with gold or platinum electrodes likely arises from
differing conditions during the growth and transfer process.

\begin{figure}
\includegraphics[width=8cm, clip]{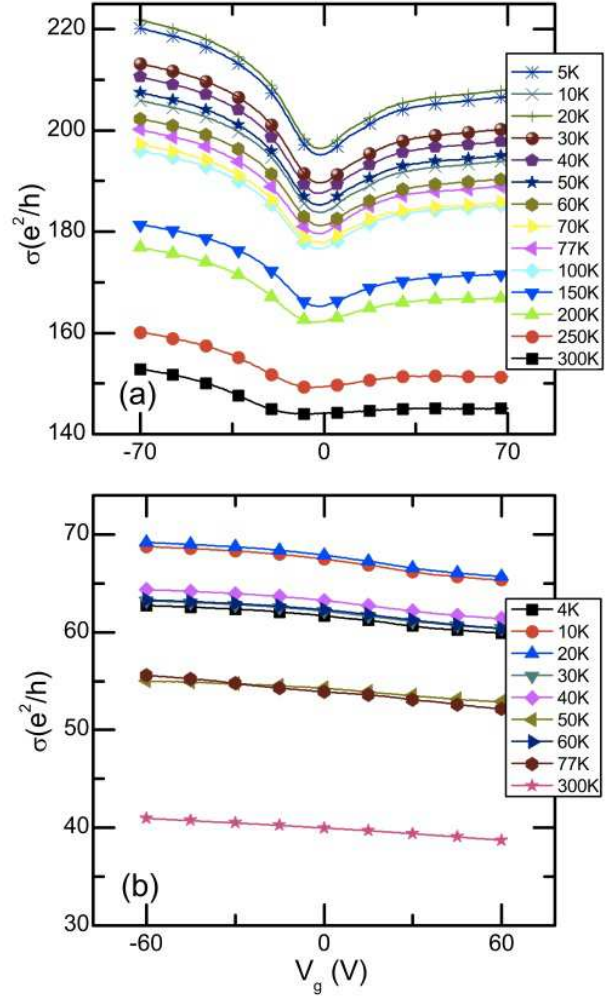}

\caption{DC transport data from (a) gold (L = 20 nm), $\mu_{20K}=4.8\times10^{2}$
cm$^{2}$/V$\cdot$s, and (b) platinum (L = 50 nm), $\mu_{20K}=3\times10^{1}$
cm$^{2}$/V$\cdot$ s, short channel electrodes. The voltage across
the electrodes was fixed at 100 mV. A clear neutrality point evolves
in the gold-based devices near $V_{g}$ = 0~V but is shifted to +$V_{g}$
in the platinum-based devices.\label{fig:fig2}}

\end{figure}

We first present data taken from our very short channel length
transistors, shown in Figure \ref{fig:fig2}. Figure \ref{fig:fig2}(a)
uses gold electrodes and Figure \ref{fig:fig2}(b) uses platinum
electrodes.  The sets of devices were prepared via identical methods,
suggesting that any contamination should be common to the two.  Both
sets of devices were annealed in vacuum for 16 hours at $100^{\circ}$C
to remove adsorbed contamination. The more aggressive traditional
method of annealing graphene devices at $400 ^{\circ}$C could not be
used for the Au electrodes in this geometry, because gold atom
mobility led to short circuits.  The more robust Pt devices were
subsequently annealed at $400 ^{\circ}$C for one hour in Ar/H$_{2}$;
minimal changes were observed from the data shown in Figure 2, as
described below.
Significant contamination on graphene requiring higher
temperature annealing is less likely to be a concern in the short channel
length transistors due to the small area probed.  

From the figure, a dramatic shift in charge NP to a positive voltage
can be seen with graphene on platinum that is not present for graphene
on gold.  This behavior was seen in 8 additional Pt devices and 9
additional Au devices. Because of differences in the chemical
potential between graphene and gold or platinum, the local charge
density will change as charges move between the metal-graphene
interface in order to equilibrate the chemical potential for electrons
across the two differing materials. This will depend on the relative
work functions of graphene and the metal contacts. Because surface
adsorbates can cause shifts in the work function of metals due to
intrinsic electric dipole moments,\cite{DeRenzi2005} it is difficult
to know the true work function of our electrodes. However, as both the
gold and platinum electrodes were handled under similar conditions, it
is likely they have similar types of contamination, leading to similar
shifts in their work function.  Previous scanning potentiometry
experiments\cite{Hamadani2006} have shown that Pt films processed in
our laboratory continue to exhibit work functions higher than
identically processed Au films, even upon limited exposure to ambient
conditions.  This difference in work functions can have notable
effects on charge injection, as has been seen in experiments on
conjugated polymers.\cite{Hamadani2006} The energy level crossover
from n- to p-type doping in graphene has been calculated to be
$\sim5.4$eV.\cite{Giovannetti2008} This then suggests the possibility
that gold electrodes will transfer electrons into graphene while
platinum electrodes will receive electrons from graphene, based on the
composite work functions of the metals and adsorbates.

The data shown in Figure \ref{fig:fig2} are consistent with charge
transfer doping of graphene, namely in the shift of the NP found in
platinum electrodes and the broadening of the $V_{g}-I_{d}$
curves.\cite{Chen2008a,McCreary2011} In the case of the platinum
electrodes, the large shift in NP to positive gate voltage suggests an
increase in hole carriers in the region between electrodes. The gold
electrodes appear to have very little doping as the NP stays centered
around 0~V. The gate voltage required to recover the NP can be used to
determine the charge density under zero gate bias, using
$n=C_{g}V_{g}/e$ . For gold electrodes, the largest gate voltage
required to reach the NP was $V_{g}=-7$V, corresponding to a charge
density of 3.2$\times$10$^{12}$ cm$^{-2}$ electrons.  For identically
processed platinum electrodes, the NP could not be reached with our
experimental setup, so the lower bound is $n=$2.72$\times$10$^{13}$
cm$^{-2}$ holes.  Subsequent annealing of the short-channel Pt
devices at 400 $^{\circ}$C, does not shift the NP to near $V_{g}=0$
in any devices.   Five Pt devices out of twenty-seven exhibited
some NP shift with this added annealing, with the NP closest to $V_{g}=0$ occurring at
$V_{g}=~38$~V, giving $n = 1.73 \times 10^{13}$~cm$^{-2}$ for that
device.  However, the remaining twenty-two Pt devices on that chip do
not show a change in NP from the data in Figure \ref{fig:fig2}.  These
data strongly suggest that platinum (with possible work function
modifying adsorbates) is a significant donor of holes into graphene,
increasing the charge carrier density by an order of magnitude over
that of gold-based devices.

\begin{figure}
\includegraphics[width=8cm,clip]{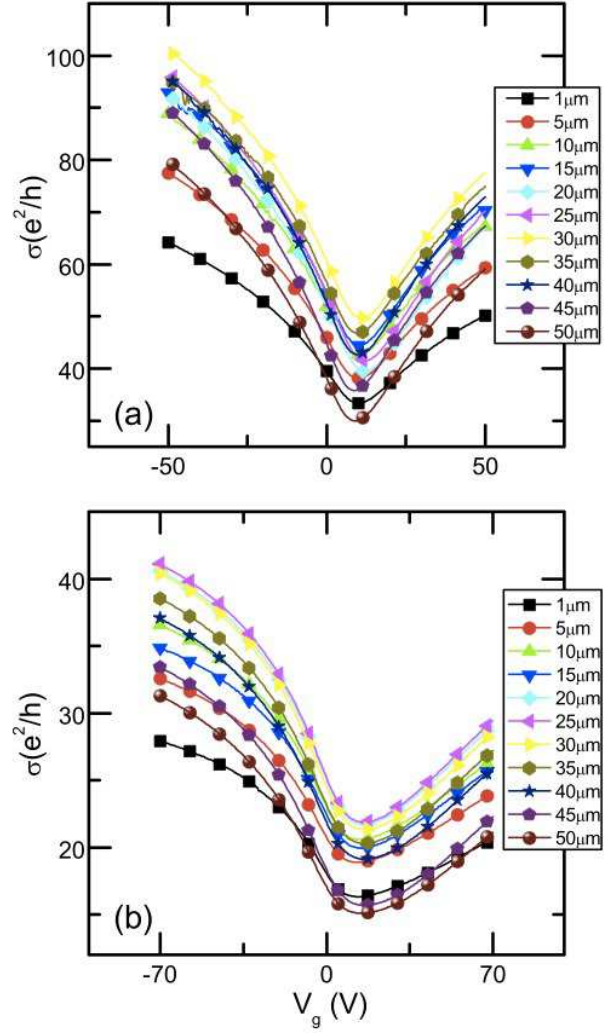}

\caption{DC transport data from (a) gold, $\mu=3.0\times10^{3}$ cm$^{2}$/V$\cdot$ s, and (b) platinum $\mu=1.0\times10^{3}$ cm$^{2}$/V$\cdot$ s, long channel (1 \textmu{}m - 50 \textmu{}m) devices, with
bias fixed at 100~mV at 300~K. No visible difference in the position
of the neutrality point is seen between the two samples.\label{fig:fig3}}

\end{figure}

To establish an upper bound on the distance that the transferred charge
travels into graphene, we measured $\sigma(V_{g})$ for our interdigitated
electrodes, shown in Figure \ref{fig:fig3}. As the channel lengths
increase for both gold and platinum devices, the influence of the
contacts on the overall charge density of the channel will decrease.
Once a maximum channel length has been exceeded, the charge density
in graphene farthest from both contacts will return to the bulk value.
Therefore, in contrast to the short-channel devices, longer channel
gold- and platinum-based electrode devices should have no NP shift
relative to each other. Initially, both types of devices showed a
large positive shift in NP toward $+V_{g}$ likely due to contamination from
the fabrication process. Both Au and Pt samples were initially
annealed under the same conditions as the devices shown in Figure
\ref{fig:fig2} ($100^{\circ}$C in vacuum for 16 hours) with no
appreciable change in NP position. The interdigitated electrodes with
graphene were then annealed at $400^{\circ}$C in an argon/hydrogen gas
mixture and annealed again at $100^{\circ}$C in vacuum for 16
hours. After the high temperature annealing, the NP in both gold and
platinum devices shifted toward 0~$V_{g}$. They each share the same
relative shift in NP for all channel lengths, likely due to additional
adsorbates not removed during the annealing process or to charged
impurities in the substrate.  Comparing the data in Figure
\ref{fig:fig3} with that in Figure \ref{fig:fig2} suggests that the charge
transfer distance between metals and our graphene is less than 500
nm (half the length of our smallest device in this geometry), consistent
with previously reported results.\cite{Mueller2009,Golizadeh-Mojarad2009}

In conclusion, we have presented electronic transport measurements
on suspended graphene transistors with a sub-100 nm channel length.
Significant shifts in NP were observed in devices made with platinum
contacts but not in devices made with gold contacts, suggesting an
increase in the local charge density in platinum devices. As the channel
length increases, the NP dissimilarity between gold and platinum electrodes
disappears. These short channel length transistors offer a method
for creating suspended graphene junctions and for creating
locally doped regions in graphene.

P.M.A. acknowledges funding support from the Office of Naval Research through the MURI programme on graphene.  D. N. and J. H. W. acknowledge support from the National Science Foundation award ECCS-0901348, and useful discussions with A. A. Fursina.


\clearpage

\clearpage

\clearpage

\end{document}